\documentclass[onecolumn,pre,superscriptaddress, longbibliography, nofootinbib]{revtex4-1}

\usepackage{hyperref}
\usepackage{url}

\usepackage{amsmath,amsfonts,amssymb}
\usepackage{graphicx}

\begin{document}

%\title{Generalized Rich-Club Ordering in Complex Networks}
%
%\author{Matteo Cinelli}
% \email{matteo.cinelli@uniroma2.it}
%
%\affiliation{ Department of Enterprise Engineering \\
%University of Rome Tor Vergata\\
%Via del Politecnico, 1 - 00133 Rome, Italy.\\
%}
%
%%Collaboration name if desired (requires use of superscriptaddress
%%option in \documentclass). \noaffiliation is required (may also be
%%used with the \author command).
%%\collaboration can be followed by \email, \homepage, \thanks as well.
%%\collaboration{}
%%\noaffiliation
%
%\date{\today}
%
%\maketitle

\begin{Large}
\centerline{\textbf{Generalized Rich-Club Ordering in Networks}}
\end{Large}

\bigskip

\centerline{Matteo Cinelli \footnote{matteo.cinelli@uniroma2.it}}

\centerline{\textit{Department of Enterprise Engineering,
‘‘Tor Vergata’’, Via del Politecnico, 1 - 00133 Rome, Italy}}

\bigskip

\noindent \textbf{Abstract} Rich-club ordering refers to the tendency of nodes with a high degree to be more interconnected than expected. In this paper we consider the concept of rich-club ordering when generalized to structural measures that differ from the node degree and to non-structural measures (i.e. to node metadata). The differences in considering rich-club ordering (RCO) with respect to both structural and non-structural measures is then discussed in terms of employed coefficients and of appropriate null models (link rewiring vs metadata reshuffling). Once a framework for the evaluation of generalized rich-club ordering (GRCO) is defined, we investigate such a phenomenon in real networks provided with node metadata. By considering different notions of node richness, we compare structural and non-structural rich-club ordering, observing how external information about the network nodes is able to validate the presence of rich-clubs in networked systems.


\section{Introduction}
\label{Intro}

Networks are characterized by a number of topological properties that are able to provide important insights into their functional aspects. Well-known examples are represented by the presence of communities \cite{newman2004finding}, i.e. subgraphs whose nodes have a higher probability to be linked to every node of the subgraph than to any other node
of the graph \cite{fortunato2016community}, or of core-periphery structures \cite{may1972will}, i.e. structures that allow for the partitioning of the network into a set of central and densely connected nodes (the core) and a set of noncentral and sparsely connected nodes (the periphery) \cite{csermely2013structure}. %When the core of a certain network is made up of hubs then such a network is said to display rich-club ordering \cite{zhou2004rich}.
When the hubs of a certain network are densely interconnected (i.e. they form a tight subgraph often referred to as core) such  a network is said to display a rich-club \cite{zhou2004rich}.
%Rich-club ordering refers to the tendency of nodes with high degree to form a core denser than expected from random \cite{colizza2006detecting}. 
%Rich-club ordering 
The presence of a rich-club is quantitatively recognized through the rich-club coefficient, called $\phi(k)$, which measures the ratio between the number of links among the nodes having degree higher than a given value $k$ and the maximum possible number of links among such nodes.
The rich-club coefficient, when compared with its expectation over a set of rewired networks with the same degree sequence of the original one, is called $\phi(k)_{norm}$ and a network is said to display rich-club ordering when $\phi(k)_{norm} > 1$.
This phenomenon has been extensively investigated \cite{cinelli2018richclub,zhou2007structural,xu2010rich,mondragon2012random,jiang2008statistical,ansell2016says} as well as recognized in several real networks \cite{zhou2004accurately,ma2015anatomy,FI2016,dedomenico2017modeling}, with special focus on neuroscience \cite{van2011rich, harriger2012rich, van2012high, collin2014structural}. Such developments have fostered further research related to the study of the network core such as its size \cite{cinelli2018richclub, ma2015richcores} and its contribution to network resilience \cite{cinelli2017resilience}, as well as its functional role \cite{grayson2014structural}.
%, both unweighted (i.e. with binary links) and weighted (i.e. with non binary-links).

Here we consider the concept of rich-club ordering by investigating the interconnections between nodes that are considered as important from a number of different perspectives. Building on this, the generalized rich-club ordering (GRCO) refers to the tendency of important nodes (under a certain declared point of view) to form a core denser than expected. The importance of nodes can be evaluated from a structural point of view, e.g. the node degree or other nodal centrality measures, and from a non-structural point of view, e.g. the node metadata. 
Node metadata refer to non-structural information, such as social or technical attributes, related to network nodes that possibly display a certain correlation with the observed network structure, and their importance is increasingly being recognized in terms of understanding networked systems \cite{newman2003mixing, park2007distribution, bianconi2009assessing, peel2017ground, hric2016network, newman2016structure}. Additionally, node metadata represent exogenous information about the network nodes (also in weighted networks) that may be impossible to split over the network links.
It follows that the study of GRCO becomes particularly interesting when dealing with networks with various node metadata; as such, we aim to investigate the interrelation between such node metadata and the network structure. 
%both structural and non-structural features.

For instance, if we consider a social network with known individuals' incomes, we may find that the nodes with the highest incomes, which are not necessarily hubs, are more interconnected than expected, while those with the highest degree are not. Moreover, it is important to recall that, despite rich-club ordering and assortativity being two related concepts, positive assortativity doesn't necessarily imply rich-club ordering, and viceversa \cite{zhou2007structural}.

\begin{figure}[htbp]
\begin{center}
\includegraphics[angle = 270, trim=0cm 0cm 0cm 0cm, clip=true, scale = 0.3]{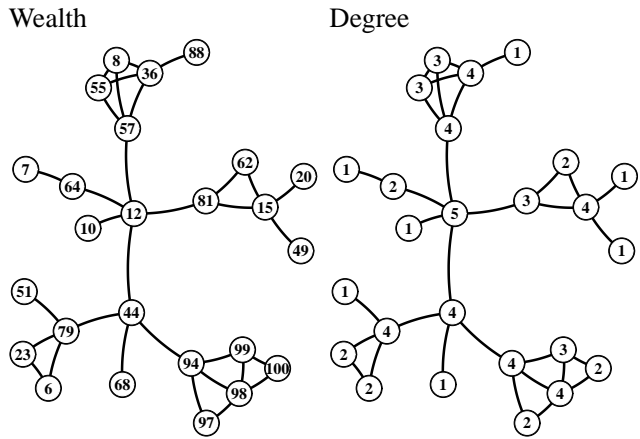}
\caption{Two toy networks with the same topology for which rich-club ordering can be evaluated with respect to structural and non-structural measures. On the left, the node labels correspond to their wealth, while on the right, the node labels correspond to their degree.}
\label{examplenet}
\end{center}
\end{figure}

In the network from Figure \ref{examplenet} we have a slightly wealth-disassortative network $r_{wealth} = -0.052$ in which, conversely, the wealthiest nodes (that are 5 if we set the wealth threshold to $w> 93$) are tightly connected (they have 7 links out of 10) despite the fact they are not the hubs of the considered network. Moreover, this network, which displays $r_{degree} =  -0.282$, doesn't show rich-club ordering (to degree) for each value of $k$ (i.e. $\phi(k)_{norm} < 1$ $\forall k$).
% Additionally, in this case, the high degree nodes are, in general, connected but not enough in order to provide evidence for rich-club ordering. 
This means that rich-club ordering to node metadata does not imply rich-club ordering to node degrees, and viceversa.
In a more general sense, we note that: in the case of rich-club ordering, in terms of node degree, it is easy to compute $\phi(k)_{norm}$, because we know which null model to use (degree-preserving rewiring \cite{maslov2002specificity})
% or its alternatives \cite{bansal2009exploring, zhou2012random, zhou2007structural, muscoloni2017rich}),
while in terms of node wealth, the situation becomes trickier since the wealth can't be directly considered a structural property. 
For this reason, in the following sections, 
%We address this issue in the following sections and 
we provide a framework as a way of determining the evaluation of GRCO together with specific null models for evaluating the significance of rich-club ordering in the case of node metadata.

\section{Related Works}
\label{RelW}

The concept of rich club ordering was initially introduced by Zhou and Mondrag{\'o}n~\cite{zhou2004rich} in order to analyze the Internet topology at the Autonomous Systems level and to provide a reasonable explanation as to why such kind of network includes tightly interconnected hubs. In order to investigate the presence of a rich-club, the authors of~\cite{zhou2004rich} introduced the rich-club coefficient $\phi(r)$ in terms of rank $r$ of the node (the sequence of ranks reflects the sequence of node degrees arranged in non-increasing order).

After the contribution of~\cite{zhou2004rich}, Colizza et al.~\cite{colizza2006detecting} considered the rich-club coefficient $\phi(k)$ in terms of degree $k$ of the node (here the sequence of the degree is opposite to the degree sequence, i.e. the node degrees are arranged in non-decreasing order). Since the work of~\cite{colizza2006detecting}, the rich-club coefficient has been mostly exploited in its $\phi(k)$ version; however, the two coefficients yield identical results.

Despite the different formulations of the rich-club coefficient, the fundamental contribution of~\cite{colizza2006detecting} derives from exploiting a null model used to detect the presence of the rich-club. Such null model exploits the procedure of degree-preserving rewiring introduced in~\cite{maslov2002specificity} in order to compute the expected rich-club coefficient $\phi(k)^{norm}$.
The reason behind the necessity of a null model for studying rich-club ordering is the observation of the monotonically increasing behavior of the rich-club coefficient $\phi(k)$.
Such a statement was subsequently denied by the authors of~\cite{zhou2007structural}, who also discussed the interrelation of the assortativity coefficient~\cite{newman2003mixing} with the rich-club structure and introduced a null model able to preserve the density of the rich-club.

Other works concerning the evaluation of rich-club ordering are related to its statistical significance~\cite{jiang2008statistical, muscoloni2017rich} in terms of p-value under different null models; to its effect on other structural measures, such as the clustering coefficient and degree assortativity~\cite{xu2010rich} that are strongly influenced by the rich-club density; to the improvement of the rich-club coefficient itself~\cite{cinelli2018richclub}, thus allowing one to consider the constraints introduced by the degree sequence of the network.

Together with its implementation on unweighted networks, the concept of rich-club ordering has been also extended to weighted networks by using various null models which are able to preserve different aspects of the network, including degree and strength distribution~\cite{serrano2008rich, opsahl2008prominence}. Other extensions of the rich-club involve dense~\cite{zlatic2009rich}, hierarchical~\cite{mcauley2007rich} and interdependent networks~\cite{valdez2013triple}.
Moreover, other contributions discuss the importance of a rich-club in network robustness~\cite{cinelli2017resilience},  spreading processes~\cite{berahmand2018effect}, as well as the generative processes and dynamics that lead to networks displaying a rich-club~\cite{mondragon2012random, csigi2017geometric}.

Finally, another important aspect related to the presence of a rich-club is to measure its size in terms of number of nodes. This has been achieved through the persistence probability of a random walker in the network~\cite{ma2015richcores} and the number of nodes that are necessary to realize a complete subgraph from the degree sequence of the considered network~\cite{cinelli2018richclub}.

\section{Evaluating rich-club ordering for the node degree}
\label{rich_coeff}

Rich-club ordering can be quantified using the coefficient $\phi(k)$:
\begin{equation}
\phi(k) = {\frac {2 E_{>k}}{N_{>k}(N_{>k}-1)}}
\label{phi}
\end{equation}
where $E_{>k}$ is the number of links among the $N_{>k}$ nodes having degree higher than a given value $k$ and $\frac {N_{>k}(N_{>k}-1)}{2}$ is the maximum possible number of links among the $N_{>k}$ nodes. Therefore, $\phi (k)$ measures the fraction of links connecting the $N_{>k}$ nodes out of the maximum number of links they might possibly share. This implies that $\phi(k) =1$ when the $N_{>k}$ nodes are arranged into a clique. When rich-club ordering is investigated, the rich-club coefficient needs to be compared against a null model in order to evaluate its significance (i.e. to test that the presence of rich-club ordering is not a natural consequence of the considered degree sequence).
The use of null models and of the normalization process of structural measures in complex networks represents a practice widely used to comprehend whether an observed pattern could have arisen by chance. For this reason, the normalization of the rich-club coefficient, suggested in~\cite{colizza2006detecting} and adopted in many further studies~\cite{bullmore2009complex, harriger2012rich, jiang2008statistical, mcauley2007rich, sporns2004organization, zamora2011exploring, zlatic2009rich}, is a necessary procedure that has to be adopted in order to take into account the significance of this index.
The normalization procedure of $\phi (k)$ involves an ensemble of rewired networks which have the same degree sequence of the one under investigation and that, if generated in a sufficiently large number, provide a null distribution of the rich-club coefficient. 
The rewiring procedure itself is simple since it chooses two arbitrary edges at each step ((a,b) and (c,d) for instance) and changes their endpoints (such that we obtain (a,d) and (c,b)) \cite{maslov2002specificity}; in cases whereby one or both of these new links already exist in the network, this step is aborted and a new pair of links is selected. The described procedure
%, which tends to a uniform sampling of random networks \cite{milo2003uniform}, 
has been widely adopted since it preserves an important network parameter represented by nodes degree; however, other procedures that aim at preserving other parameters may be adopted \cite{bansal2009exploring, mondragon2012random, zhou2007structural, mondragon2014network,muscoloni2017rich}.

In general, the normalized rich-club coefficient~\cite{colizza2006detecting} is defined as: 
\begin{equation}
\phi(k)_{norm} = {\frac {\phi(k)}{\phi(k)_{rand}}}
\label{phinorm}
\end{equation}
where $\phi(k)_{rand}$ is the average rich-club coefficient across the set of rewired networks (typically 1000 networks \cite{collin2014structural,van2012high,dedomenico2017modeling}) and we observe rich-club ordering when $\phi(k)_{norm} > 1$.

Additionally, in \cite{zhou2007structural} it is argued that when the considered network is made up of nodes whose maximum degree $k_{max}$ is larger than the cut-off degree $k_s$ \cite{boguna2004cut} (i.e. the quantity for which it is impossible to obtain networks with no degree-degree correlation) the degree-preserving rewiring could produce randomized networks with a rich-club coefficient that is too close to the initial one. This is because the rewiring procedure, in which couples of links are uniformly sampled, could cause the disruption of several high-degree to low-degree connections with the consequent creation of high-degree to high-degree connections. Indeed, since a high proportion of links is attached to hubs the probability of picking a couple of links whose endpoints are hubs is relatively high, particularly when there are nodes with degree $k_{max}>k_s$.

\section{Null Models for the evaluation of rich-club ordering}

The evaluation of rich-club ordering in the case of degree exploits a null model that rewires the network while keeping its degree sequence. As the degree can be considered a structural attribute of the node, a null model that evaluates different network topologies (i.e. alters the original network structure while keeping certain fundamental properties) constitutes a reasonable choice. The same choice seems to be reasonable also in the case of other structural properties of the node (such as centrality measures) even if the degree-preserving rewiring doesn't keep the same value of centrality over the nodes due to the topology being subject to change.

In the case of non-structural attributes (i.e. node metadata), the structural rewiring doesn't seem to be the unique option. Indeed, we may be interested in knowing if different arrangements of the node metadata over the same network structure are able to unveil rich-club ordering as well. In other words, we may also be interested in using a null model that keeps the original network structure while reshuffling the node metadata. 

More intuitively, when we evaluate rich-club ordering with link rewiring we are basically asking the question: does the considered network possess a topology so unusual that it allows room for rich-club ordering?  
Alternatively, if we evaluate rich-club ordering with metadata reshuffling we are basically asking the question: does the considered network possess an arrangement of node metadata so unusual that it allows room for rich-club ordering?

As an example, let us suppose that a relatively large clique of wealthy nodes is present in a certain social network, like that in the example of Section \ref{Intro}. The two questions from above then become: 
%\begin{enumerate}
%\item is it such particular that this clique of wealthy nodes exists?
%\item is it such particular that the wealthy nodes are arranged into a clique?
%\end{enumerate}
\begin{enumerate}
\item Is it so peculiar, given the degree of the wealthy nodes, to observe a realization that contains a clique made up of such nodes?
\item Is it so peculiar, given the network structure, to observe a distribution of the node metadata such that the wealthy nodes are arranged into a clique?
\end{enumerate}
Consequently, we may be interested in understanding if the metadata distribution is related to the presence of structural rich-club ordering, and if rich-club ordering, evaluated with respect to node metadata, can be interpreted as a reinforcement (or a weakening) of the evidence of structural rich-club ordering. Indeed, the exploitation of node metadata takes into account an additional layer of information that derives from the coupling between the network structure and the node metadata. 

Thus, in order to evaluate rich-club ordering in the case of node metadata we suggest a comparison between the number of links observed among the rich nodes and the average number of links observed among such rich nodes over two different ensembles: the former made up of networks and obtained via the rewiring of links; the latter made up of vectors of metadata and obtained via the reshuffling of node attributes. 

These two methods generate different ensembles into which different aspects of the original network are kept. 
In the first case (link rewiring) we lose the original network topology while we keep its degree sequence and its degree-attribute correlation. In the second case we lose the degree-attribute correlation but we keep the original network structure.
Both the methods, despite their clear differences, seem to provide a valid basis of comparison in the case of node metadata. 
Moreover, when we evaluate GRCO considering node metadata, if the node metadata and the node degrees don't display a significant correlation (either positive or negative), we could observe a set of rich nodes with very heterogeneous degrees.
%very heterogeneous degree profile of the rich nodes. 
%we do not  have a significant correlation (either positive or negative) among the node metadata and the node degree we could have a very heterogeneous degree profile of the rich-nodes. 
This doesn't imply, however, the absence of a dense subgraph made up of rich nodes. As an example, if we suppose that a subgraph of 15 nodes has the highest metadata values, then the degree of such nodes has to be at least 14 in order to realize a clique that is connected to the rest of the graph; thus, considering a sufficiently large network, such nodes don't have to be necessarily hubs in order to establish a rich-club from the metadata point of view. It follows that a positive correlation between node metadata and node degree has an effect on the rich-club evaluation and that, in certain cases, the presence of rich-club ordering with respect to node degree could also indicate rich-club ordering with respect to node metadata. 

As an extreme case, if we have perfect positive correlation among the considered attribute and the node degree $\rho_{x,k} = 1$, the sorting of the node metadata will correspond to the sorting of the node degrees (i.e. to the degree sequence). Therefore, we will evaluate rich-club ordering with the same sorting of nodes but with implications that will differ depending on the null model that we choose to adopt.
%a different meaning given by the null model that we choose to adopt.
%THIS IS INTERESTING BECAUSE THE ENSEMBLE OF REWIRED NETWORKS MAY HAVE A DIFFERENT NUMBER OF ELEMENTS THAN TH ENSEMBLE OF RESHUFFLED NETWORKS.
Additionally, we should consider that the degree-attribute correlation, when significant, represents an important feature of the considered network which, if not completely dropped, represents a remarkable element to further stress the presence of rich-club ordering. Therefore, rich-club ordering in the case of node metadata could be evaluated with respect to random reshuffling, and this would provide us an ensemble of reshuffled labels that would be, in general, uncorrelated with the network structure in terms of node degrees, but also stressed with respect to a reshuffling procedure that, similarly as in \cite{laniado2016gender},  keeps the distribution of the degree-attribute correlation somewhat closer to that of the original network. 

%\textbf{This leads us to the following question: 
%is a random shuffle of node metadata a sufficient null model for the evaluation of rich-club ordering? Maybe not. Indeed, the link rewiring algorithm (as well as many other null models) has been introduced in order to provide a null model that, as we discussed above, keeps some structural aspects of the considered network.
%ARE THE TOPOLOGIES CORRELATED? HOW TO MEASURE THE CORRELATION AMONG DIFFERENT TOPOLOGIES? HOW TO MEASURE THE STRUCTURAL SIMILIRATY?
%A random node attribute shuffling would provide us an ensemble of reshuffled labels that would be somehow uncorrelated with the network structure in terms of node degrees. Therefore, in the case we observe a significant correlation between the node attributes and the node degrees we might be interested in adopting a null model that keeps the distribution of the degree-attribute correlation somehow closer to the one of the original network.}
%This might be especially valuable in the analysis of empirical networks since the evaluation of certain phenomena is conducted over a finite set of networks whose links have been rewired a finite number of times. 
In order to address this point,  in Section \ref{alternative} we introduce a procedure which, by keeping a certain node-attribute correlation, aims at further stressing the presence of rich-club ordering by comparing it with a somewhat unfavorable set of metadata shuffles.
%MOREOVER WHEN YOU HAVE A NULL MODEL, THE MORE YOU KEEP (OTHER THAN THE PHENOMENON) THE BETTER

\section{Evaluating rich-club ordering for structural measures different from degree}

When evaluating rich-club ordering (i.e. to compute $\phi_{norm}$) with respect to structural measures different from node degree, we should take into account that the degree-preserving rewiring entails the two following aspects:
\begin{enumerate}
\item The structural measure of node $i$ may change its value due to rewiring.
\item The number of nodes that retain a value of the considered structural measure above the threshold for which we evaluate rich-club ordering may change.
\end{enumerate} 
In order to address this problem we evaluate rich-club ordering by creating a ranking of nodes; in other words, we consider the rich-club coefficient as a measure of position. We thus rank, for each network in the random ensemble, the nodes in non-decreasing order of the considered measure and we assign each of them to a position $p$ with $p \in [1,N]$. Then we consider the number of links among the nodes that have a rank greater than a given value $p$.
In other words, while the degree sequence is fixed across rewired networks, the rewiring procedure may alter the values of the structural measure associated to each node and consequently the number of nodes with a certain value of a such measure. In order to address this issue and consider the same amount of nodes at each iteration (which corresponds to keeping the denominator of $\phi_i$ constant for a certain $i$) both in the original network and in the randomized ensemble, we evaluate rich-club ordering by creating a ranking of such nodes. Therefore, the nodes of the original network and of its randomized instances are ranked in non-decreasing order of the considered structural measure and assigned with a position $p \in [1,N]$. In such a way, for each network, the node with the lowest value of the considered measure will be in position 1 while that with the highest value will be in position $N$, despite the possible differences of highest/lowest values among different networks.

%Therefore, in order to compute $\phi(p)$ we check, for each value of $p$, the density of connections among nodes retaining a value of the considered structural measure higher than that in position $p$ and:
Therefore, in order to compute $\phi(p)$ we compute the density of connections among nodes whose index of position is greater than $p$:
\begin{equation}
\phi(p) = {\frac {2 E_{>p}}{N_{>p}(N_{>p}-1)}}
\label{phinormcentr}
\end{equation}
where $E_{>p}$ is the number of edges among the $N_{>p}$ nodes with centrality value greater then the value in position $p$
and $\frac {N_{>p}(N_{>p}-1)}{2}$ is the maximum possible number of edges among the $N_{>p}$ nodes.

By using this procedure we obtain $\phi(p)_{norm} = \frac{\phi(p)}{\phi(p)_{rand}}$ where $\phi(p)_{rand}$ is the average of $\phi(p)$ over the random ensemble. It is worth mentioning that this way of computing the coefficient $\phi(p)$ (as a measure of position) is similar to that proposed in the paper that originally discussed rich-club ordering \cite{zhou2004rich}. Additionally, this measure is also related to the rich-club coefficient for weighted networks (i.e. networks with non-binary links) proposed by \cite{opsahl2008prominence}. In \cite{opsahl2008prominence}, indeed, they consider structural measures such as the node strength (i.e. the sum of the weights attached to the links of a certain node) and the average weight (i.e. the ratio between the node strength and the node degree), and they normalize the rich-club coefficient with a method that reshuffles the weights over the links and then links themselves. %In this case, however, the value of the node attribute is obtained by manipulating (for instance summing up or normalizing) all the values of links attributes and therefore, given the correlation of the considered attribute with the structure, the randomization of node attributes would have no sense.

\section{Evaluating rich-club ordering for non-structural measures}

The rich-club coefficient in the case of node metadata can be computed only for scalar metadata as we need a quantity which, like the node degree or other structural measures, can be sorted in a certain order.
The coefficient can be easily derived from the case of node degree by considering, instead of the degree $k$, a certain value $m$ corresponding to the value of the node metadata.
Therefore, rich-club ordering can be discovered via the coefficient $\phi(m)$:
\begin{equation}
\phi(m) = {\frac {2 E_{>m}}{N_{>m}(N_{>m}-1)}}
\label{phinormmeta}
\end{equation}
where $E_{>m}$ is the number of edges among the $N_{>m}$ nodes having metadata value higher than a given value $m$ and $\frac {N_{>m}(N_{>m}-1)}{2}$ is the maximum possible number of edges among the $N_{>m}$ nodes.

The normalized rich-club coefficient, $\phi(m)_{norm}$, can be derived by considering $m$ as the value corresponding to a certain value of the node metadata, whilst considering $\phi(m)_{rand}$ from two different perspectives.
In other words, in the case of node metadata, we obtain two values of $\phi(m)_{rand}$ that depend on the null model that we use.
% Moreover, in the case of non-structural measures both the rewiring and reshuffling procedures do not obviously affect the values within the metadata vector which remains the same.
In the case of link rewiring, $\phi(m)_{rand}$ is called $\phi(m)_{rand}^{rew}$ and we use, as for example in \cite{almeira2017structure}, the coefficient:
\begin{equation}
\phi(m)_{norm}^{rew} = \frac{\phi(m)}{\phi(m)_{rand}^{rew}}
\label{phinormmetarew}
\end{equation}
while in the case of metadata reshuffling, $\phi(m)_{rand}$ is called $\phi(m)_{rand}^{resh}$ and we use the coefficient:
\begin{equation}
\phi(m)_{norm}^{resh} = \frac{\phi(m)}{\phi(m)_{rand}^{resh}}
\label{phinormmetaresh}
\end{equation}
Finally it is worth adding that, in the case of non-structural measures, both the rewiring and reshuffling procedures do not obviously affect the values of the metadata vector whose entries, in the latter case, are only modified in terms of position.

\section{A framework for the evaluation of Generalized Rich-Club Ordering (GRCO)}

In Table~\ref{frameworktab} we propose a framework for the evaluation of generalized rich-club ordering (GRCO~\footnote{R code for the evaluation of  GRCO exploits the libraries igraph~\cite{csardi2006igraph} and brainGraph~\cite{watson2018brain} and is available at  \url{https://github.com/cinHELLi}}).
%\squeezetable
\begin{table}[h!]
\centering
\begin{tabular}{|r|c|}
 \hline
  \multicolumn{2}{|c|}{Consider a certain node feature. It can be:} \\
  \hline
  \hline
  1 & Structural \\
  2 & Non-structural \\% \cline{2-2}
%  11111000000 & binary \\
  \hline \hline
   \multicolumn{2}{|c|}{if Structural, it can be:} \\
  \hline
   degree & Compute $\phi(k)_{norm}$ with degree-preserving rewiring \\
   & as in Equation \ref{phinorm} \\
  $\neq$ degree &  Compute $\phi(p)_{norm}$ with degree-preserving rewiring \\
  & as in Equation \ref{phinormcentr} \\
   \hline \hline
   \multicolumn{2}{|c|}{if Non-Structural:} \\
  \hline
   \multicolumn{2}{|c|}{Compute $\phi(m)_{norm}^{rew}$ with degree-preserving rewiring} \\%(i.e. $\phi(m)_{norm}^{rew}$) 
    \multicolumn{2}{|c|}{as in Equation \ref{phinormmetarew}} \\
    \multicolumn{2}{|c|}{Compute $\phi(m)_{norm}^{resh}$ with metadata reshuffling}\\% (i.e. $\phi(m)_{norm}^{resh}$) 
    \multicolumn{2}{|c|}{as in Equation \ref{phinormmetaresh}} \\
   \hline
\end{tabular}
%\vspace*{-4mm}
\label{frameworktab}
\caption{Generalized framework for the evaluation of rich-club ordering}
\end{table}

\section{Application}
\subsection{Social Network}
We test the introduced framework in the case of a criminal social network \cite{grund2015ethnic}. The network is made up of the relationships ($m = 315$ that we consider unweighted) among confirmed members ($n = 54$) of a London street gang between 2005-2009. We choose this network as it comes with various node metadata (which are an important piece of information in criminal networks \cite{villani2018avirtuous}) such as age, number of arrests and convictions. We compute the rich-club coefficient for two structural characteristics of nodes, degree and eigenvector centrality, and for three non-structural characteristics, corresponding to the node metadata using degree-preserving rewiring ($k_s \simeq k_{max} = 25$) and node metadata reshuffling.
\begin{figure}[h]
\begin{center}
\includegraphics[angle = 270, trim=0cm 0cm 0cm 0cm, clip=true, scale = 0.4]{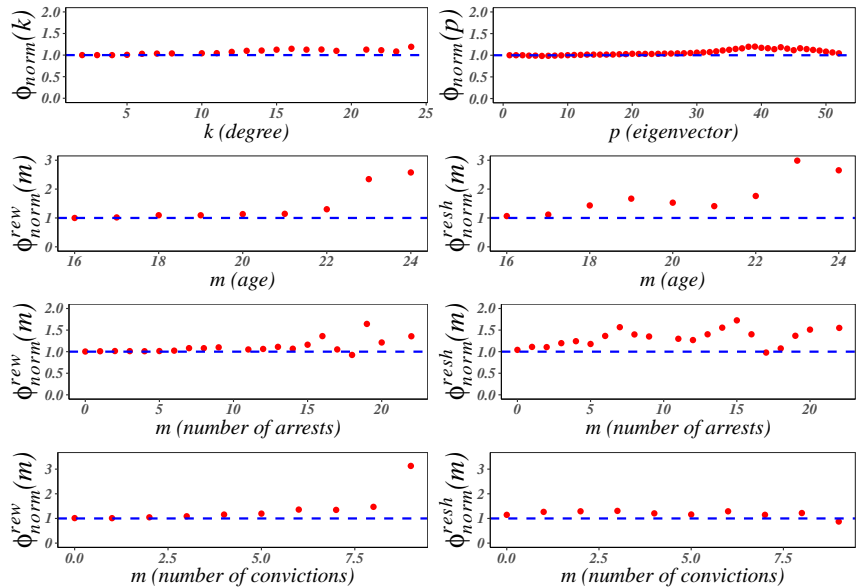}
\caption{Curves of the coefficient $\phi_{norm}$ for the criminal social network. The dashed line occurs in correspondence with $\phi_{norm} = 1$, the threshold above which we observe rich-club ordering.
From top-left we compute GRCO for: degree, eigenvector, age (rewiring), age (reshuffling), arrests (rewiring), arrests (reshuffling), convictions (rewiring), convictions (reshuffling).}
\label{criminals_rich}
\end{center}
\end{figure}
In Figure \ref{criminals_rich}, we observe that the considered network displays rich-club ordering $\phi(k)_{norm} > 1$ to degree. Thus, in such a network, hubs happens to be more connected than what we observe, on average, across the rewired network ensemble. The network displays also what we can call power-club ordering, as the nodes with highest eigenvector centrality are also tightly connected. The latter result is, however, expected since the degree and the eigenvector centrality are, in general, positively correlated \cite{valente2008correlated}. 

When we consider the node age and the number of arrests we also observe rich-club ordering. We especially observe how, in the two cases, the metadata reshuffling entails a stronger rich-club ordering $\phi(m)_{norm}^{resh} \geq \phi(m)_{norm}^{rew} > 1$ than the link rewiring. This means that the metadata are arranged in a way that elicits rich-club ordering and that this arrangement is hard to replicate via random label reshuffling. The fact that the two measures are both in favour of rich-club ordering denotes that the presence of this phenomenon is far from being random from different perspectives, thus underlining the importance of the interplay among the node metadata and the network topology. 
Conversely, we observe slightly discordant results when the number of convictions is taken into account. For a certain value of $k$ ($k = 9$), rich-club ordering appears to be absent from the structural point of view and present from the metadata point of view. %This may be due to the fact that nodes with the highest number of convictions ($>9$ in this case) also have low degree. Indeed, there are 4 nodes with a number of convictions greater than 9, and they have degree $d = [2,2,14,16]$ (this also confirms the possibility of a heterogeneous degree-profile of nodes in the case of rich-club ordering with respect to node metadata). The observation of a discordant result is also explained by the low value of the correlation coefficient between the degree and the number of convictions, which is $\rho_{deg, conv} = 0.058$.    
Such a discrepancy is due to the fact that nodes with the highest number of convictions ($>9$ in this case) also have heteregeneous degrees. Indeed, there are five nodes with more than nine convictions, and they have degree $d = [2,2,2,14,16]$. The maximum number of links that those five nodes could share is ten, while in the actual network they only share two links. Such a small amount of links is also due to the presence of nodes with very low degree. Moreover, since about $90\%$ of the nodes in the actual network have degree higher than 2 and the network, being relatively dense, displays several complete subgraphs of size 5, then we should expect reshuffled instances displaying a rich-club coefficient below one. In other words, we should expect a higher number of links among highly convicted nodes in randomized networks. In this case, the value of the rich-club coefficient $\phi(k)_{norm}^{resh}$ indicates that highly convicted elements tend to avoid each other.

More technically, this analysis entails that the degree heterogeneity of the nodes that we take into account, when related to other elements such as the network density, is able to explain the discrepancies observed between the coefficients $\phi(k)_{norm}^{rew}$ and $\phi(k)_{norm}^{resh}$. A discordant result between the two coefficients is also partially explained by the low value of the correlation coefficient between the degree and the number of convictions, which is $\rho_{deg, conv} = 0.058$.

\subsection{Linguistic Network}

We consider the global language network in which each node represents a language and links connect languages that are likely to be co-spoken~\cite{ronen2014links}. In more detail, languages are connected according to the frequency of book translations, i.e. two languages are connected if, at the very least, a book is translated from one language to the other. The data are pre-processed in order to consider only the largest connected component of the language network compatibly with the availability of node metadata.
The resulting network has $n=54$ nodes and $m = 104$ links, and the node metadata are represented by two elements: the GDP (gross domestic product) per capita for a language and the number of speakers of a certain language. As described in~\cite{ronen2014links} the GDP per capita for a language is measured as the average contribution of a single speaker of language $l$ to the world GDP, and is calculated by adding the contributions of speakers of $l$ to the GDP of every country, and dividing the sum by the number of speakers of $l$. The number of speakers of a certain language are computed using the speaker estimates from the June 14, 2012 version of the Wikipedia Statistics page, as explained in~\cite{ronen2014links}.
%FURTHER DETAILS ABOUT THE DATA SET
%
%(measured as the average contribution of a single speaker of language l to the world GDP, and is calculated by summing the contributions of speakers of l to the GDP of every country, and dividing the sum by the number of speakers of l)
%
%We use language speaker estimates from the June 14, 2012 version of Wikipedia
%Statistics page (17). These estimates include all speakers of a language, native and nonnative
%alike.
%
\begin{figure}[h]
\begin{center}
\includegraphics[angle = 270,trim=0cm 0cm 0cm 0cm, clip=true, scale = 0.4]{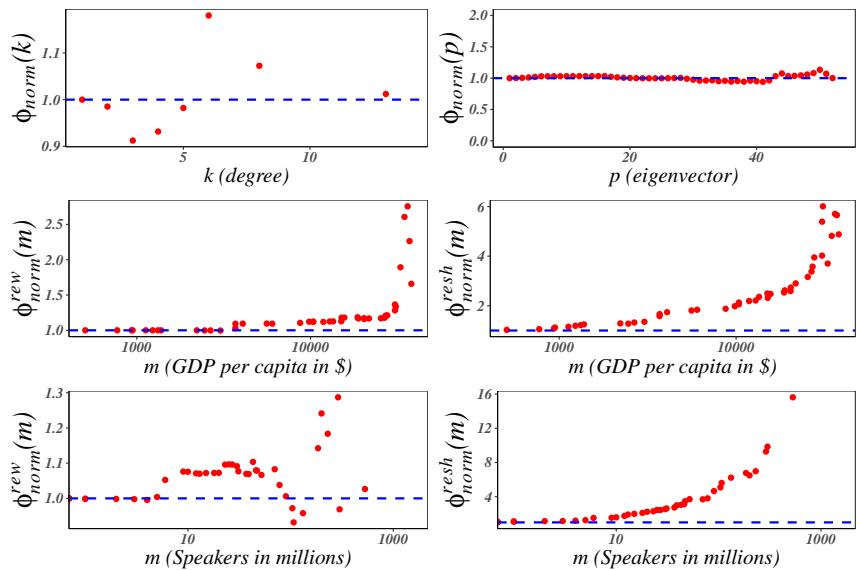}
\caption{Curves of the coefficient $\phi_{norm}$ for the global language network. The dashed line occurs in correspondence with $\phi_{norm} = 1$, the threshold above which we observe rich-club ordering.
From top-left we compute GRCO for: degree, eigenvector, GDP per capita (rewiring), GDP per capita (reshuffling), amount of speakers (rewiring), amount of speakers (reshuffling).}
\label{lang_rich}
\end{center}
\end{figure}
In Figure~\ref{lang_rich} we observe how the global language network tends to display rich-club ordering from both a structural and non-structural point of view. It is interesting how languages with a relatively high number of speakers are, in some instances, less interconnected than expected, meaning that certain books written originally in a widely spoken language were not directly translated into the other most spoken languages. It is likely that such books were first translated into other less spoken languages, which acted as mediums.
 
\subsection{Transportation Network}
\begin{figure}[h]
\begin{center}
\includegraphics[angle = 270,trim=0cm 0cm 0cm 0cm, clip=true, scale = 0.4]{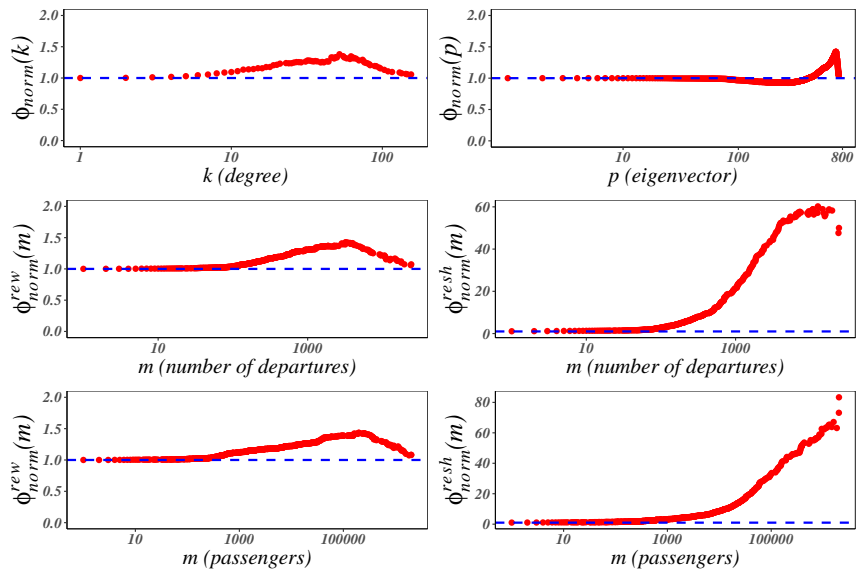}
\caption{Curves of the coefficient $\phi_{norm}$ for the US airports network. The dashed line occurs in correspondence with $\phi_{norm} = 1$, the threshold above which we observe rich-club ordering.
From top-left we compute GRCO for: degree, eigenvector, number of departures (rewiring), number of departures (reshuffling), number of passengers leaving (rewiring), number of passengers leaving (reshuffling).}
\label{airport_rich}
\end{center}
\end{figure}

We also consider the case of US airports network of domestic flights in December 2010 (the network is considered in its undirected/unweighted version with $n=745$, $m=4618$ and $k_s < k_{max} = 166$) \footnote{The network is available in the \textit{igraphdata} package for R~\cite{R}} in which the number of flights departing from each airport and the number of passengers leaving a certain airport are used as node metadata. These two quantities (reasonably) show a very high correlation with the node degrees $\rho(pass, deg) = 0.906$ and $\rho(dep, deg) = 0.928 $. For this reason, when we evaluate rich-club ordering with respect to node metadata, we observe both positive but very different values of $\phi(m)_{norm}^{rew}$ and $\phi(m)_{norm}^{resh}$. In more detail, in Figure \ref{airport_rich}, we observe very high values of $\phi(m)_{norm}^{resh}$ that depend on the fact that a random reshuffling of the node metadata causes a complete loss of the observed correlation between the metadata and the degree. The US airports network displays rich-club ordering to degree, to eigenvector centrality (from a certain point) and to the metadata values. In the latter case, we observe that, because of the random reshuffling, the obtained values are clearly on a different scale than those obtained in the case of link rewiring. In other words, the number of links among nodes with the highest metadata values is far greater than that expected by chance. This implies that the arrangement of node metadata deriving from the original network is significant and difficult to replicate (because of the degree-attribute correlation) using the current null model (i.e. a model that randomly redistributes the node metadata). Moreover, this result confirms the presence and the significance of interconnections among important airports from a wide array of perspectives connected to both the traffic generated by the airports, as well as the airports themselves.

\subsection{Technological Network}

We consider a network obtained from the data of the seventh Framework Programme for Research and Technological Development (FP7), provided by the European Commission (EC).
%FP7 (2007-2013) is a funding program created by the European Commission. 
%FP7 had the strategic objectives of strengthen the scientific and technological base of European industry and encourage its international competitiveness. 
FP7 was run from 2007 until 2013 with a total budget of over €50 billions. Most of the budget was spent on grants to both European and global research institutions to co-finance research, technological development and demonstration projects.
%
%FP7 was run from 2007 until 2013 with a total budget allocated of over 50 billions of euros. The most part of the budget was spent on grants to research institutions all over Europe and beyond to co-finance research, technological development and demonstration projects. %The grants have been assigned on the basis of calls for proposals and a peer review process.running research infrastructures, organisations and researchers from third countries, international organisations, civil society organisations.
Among the different lines of funding of FP7, we consider the data of projects related to the call for environmental issues.

Using such data we first build a bipartite network in which one partition is made up of projects while the other is made up of participants of projects. A link between the partitions exists if an institution participated in project. Then we perform a one-mode projection of the bipartite network in a way such that two institutions are connected if they participated in the same project. The resulting network has $n=2739$ and $m=45667$ and we consider as node metadata the contribution of the EC to each institution, measured in euros.

\begin{figure}[h]
\begin{center}
\includegraphics[angle = 270,trim=0cm 0cm 0cm 0cm, clip=true, scale = 0.4]{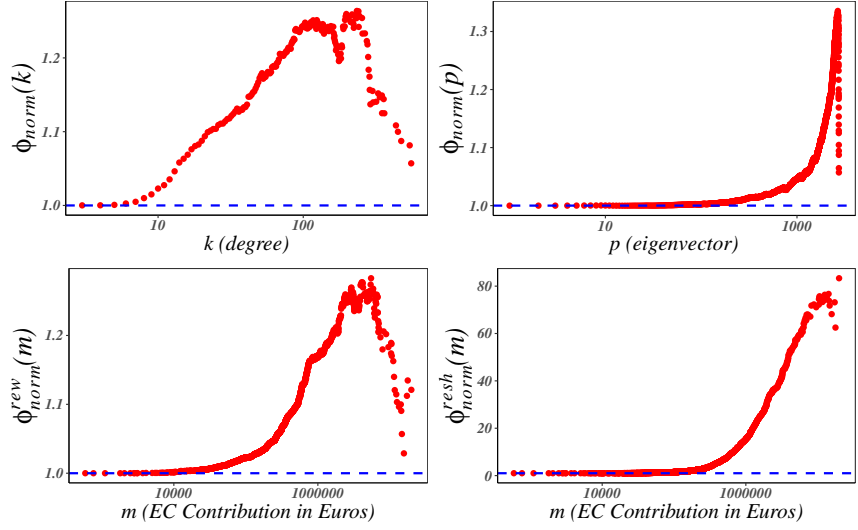}
\caption{Curves of the coefficient $\phi_{norm}$ for the FP7 projects network. The dashed line occurs in correspondence with $\phi_{norm} = 1$, the threshold above which we observe rich-club ordering.
From top-left we compute GRCO for: degree, eigenvector, European Commission contribution (rewiring), European Commission contribution (reshuffling).}
\label{fp7_rich}
\end{center}
\end{figure}

The network of institutions that we take into account has a very peculiar structure in that the participants in each project are connected in a complete subgraph, while institutions participating in multiple projects connect such dense substructures. This network, made up of several interconnected cliques, is particularly apt to being studied in terms of rich-club ordering to node metadata due to its particular structure. Indeed, we can foresee how the rewiring procedure would break up the multiple cliques, which represent each financed project and thus provide evidence for structural rich-club ordering.
When we analyze rich-club ordering in terms of contribution of the EC, i.e. when we ask ourselves if the richest nodes (in terms of received funds) are arranged into a rich-club , the reshuffling procedure represents a more suitable null model since it preserves the network structure while changing the degree-metadata correlation. 
In Figure~\ref{fp7_rich}, we observe that the considered network displays rich-club ordering from both a structural and non-structural point of view, confirming, as also observed in~\cite{ma2015anatomy} for projects funded by the Engineering
and Physical Sciences Research Council of United Kingdom, how research funds are allocated to rich-clubs. The fact that an elite circle
of academic institutions tends to over-attract funding~\cite{szell2015research} represents a major problem in research that needs to be investigated in other datasets and addressed with proper measures and interventions aimed at reducing evident inequalities.
%
%Nodes: Institutions
%Links: exist if two nodes participate to the same project.
%Node Metadata: Contribution of European Commission (Amount of Funds €)
%Are the richest (in €) nodes arranged into a rich-club?
%In this case metadata shuffling seems to be most appropriate choice.
%Such network is made up of multiple cliques (each project is a clique)
%In this case link rewiring would not preserve this very peculiar structure
%
\section{An alternative to random reshuffling}
\label{alternative}

Considering the reasoning outlined above, we suggest a procedure that, based on a certain parameter, is able to reshuffle the node metadata while keeping a degree-metadata correlation profile closer to that of the original network. Indeed, we aim at investigating GRCO discerning between two cases: where rich-club ordering is discovered due to a distribution of the node metadata that is significant with respect to an appropriate null model for the considered case and where rich-club ordering is discovered due to the comparison against networks whose attributes distribution is too far from the original one.
%In other words, we aim at discovering something that doesn't depends on the weaknesses of the null model (i.e. too much randomness because of the degree-attribute correlation disruption) and to further stress the presence of rich-club ordering.

\begin{figure}[h]
\begin{center}
\includegraphics[trim=0cm 0cm 0cm 0cm, clip=true,keepaspectratio, scale = 0.4]{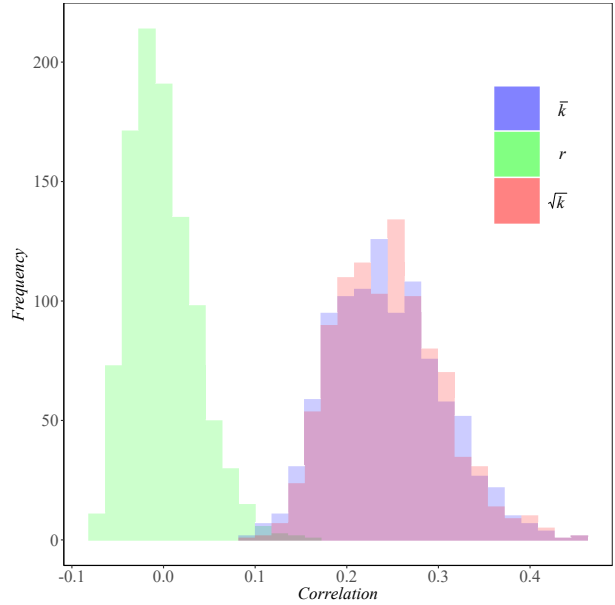}
\caption{Histograms displaying the frequencies of correlation values computed using 1000 shuffled vectors of node metadata. We choose 1000 shuffled vectors as we also consider 1000 rewired networks when computing the normalized rich-club coefficient in the case of structural measures. In the legend, $r$ refers to the random mixing of the node metadata while $\overline{k}$ and $\sqrt{k}$ refer to the mixing parameters of the proposed procedure of metadata shuffling.}
\label{histograms}
\end{center}
\end{figure}

The procedure is based on the idea of swapping a couple of metadata values whose corresponding entries in the metadata vector are at a certain distance $s$ from one another, and it is made up of the following steps:
\begin{enumerate}
\item Consider the vector of metadata of length $N$ and choose randomly an entry in position $i \in [1,N]$
\item Select the parameter $s \in [1,N]$ which determines the range of the metadata swap. 
In other words, $s$ determines the distance of the randomly chosen entry, in position $i$, from the candidate entry, in position $i' = i \pm s$, which will be selected for the swap
\item Select the direction, $\delta \in \{ 0,1 \}$, of the swap with a Bernoulli trial with probability $p = 0.5$

If $\delta = 0$ set $i' = i-s$

If $\delta = 1$ set $i' = i+s$ 
\item If $i - s < 1$ and $\delta = 0$ there is no available entry, in position $i'$, for the swap. Thus, pick uniformly at random one entry in position $[1, i-1]$ if $i \neq 1$, or in position $i' = 1$ if $i = 1$. 
Swap the entries in position $i$ and $i'$
\item If $i+s > N$ and $\delta = 1$ there is no available entry, in position $i'$, for the swap. Thus, pick uniformly at random one entry in position $[i+1,N]$ if $i \neq N$, or in position $i' =N$ if $i = N$. 
Swap the entries in position $i$ and $i'$
\item Else swap the entries in position $i$ with the entry in position $i'$ where $i' = i \pm s$ depending on the value of $\delta$
\item Repeat the steps from 3 to 6 $O(M)$ times
\end{enumerate}
\begin{figure}[h]
\begin{center}
\includegraphics[trim=2cm 2cm 0cm 2cm, clip=true,keepaspectratio,scale = 0.35]{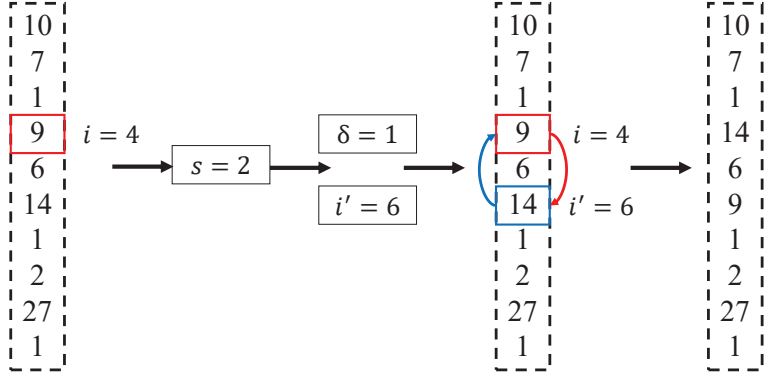}
\caption{Example iteration of the procedure of metadata reshuffling on a random metadata vector. A entry in position $i=4$ is randomly selected and the parameter $s$ is set to s=2. We suppose that the result of the Bernoulli trial is $\delta = 1$ and thus $i'=i+s=6$. Finally the entries in position $i$ and $i'$ are switched and another iteration is repeated using the new metadata vector.}
\label{iteration}
\end{center}
\end{figure}
In Figure~\ref{iteration} we pictorially display an iteration of the proposed procedure while in Figure~\ref{histograms} we show three distributions of degree-attribute correlation in the case of the US airports networks, considering as node metadata the passengers leaving each  airport (node). The three represented cases are: random reshuffling; reshuffling with the described procedure using as a parameter the mean degree $s= \overline{k}$; reshuffling with the described procedure using as a parameter the square root of the degree of the selected node $s = \sqrt{k_i}$.  

\begin{figure}[h]
\begin{center}
\includegraphics[angle = 270, trim=0cm 0cm 0cm 0cm, clip=true, scale = 0.4]{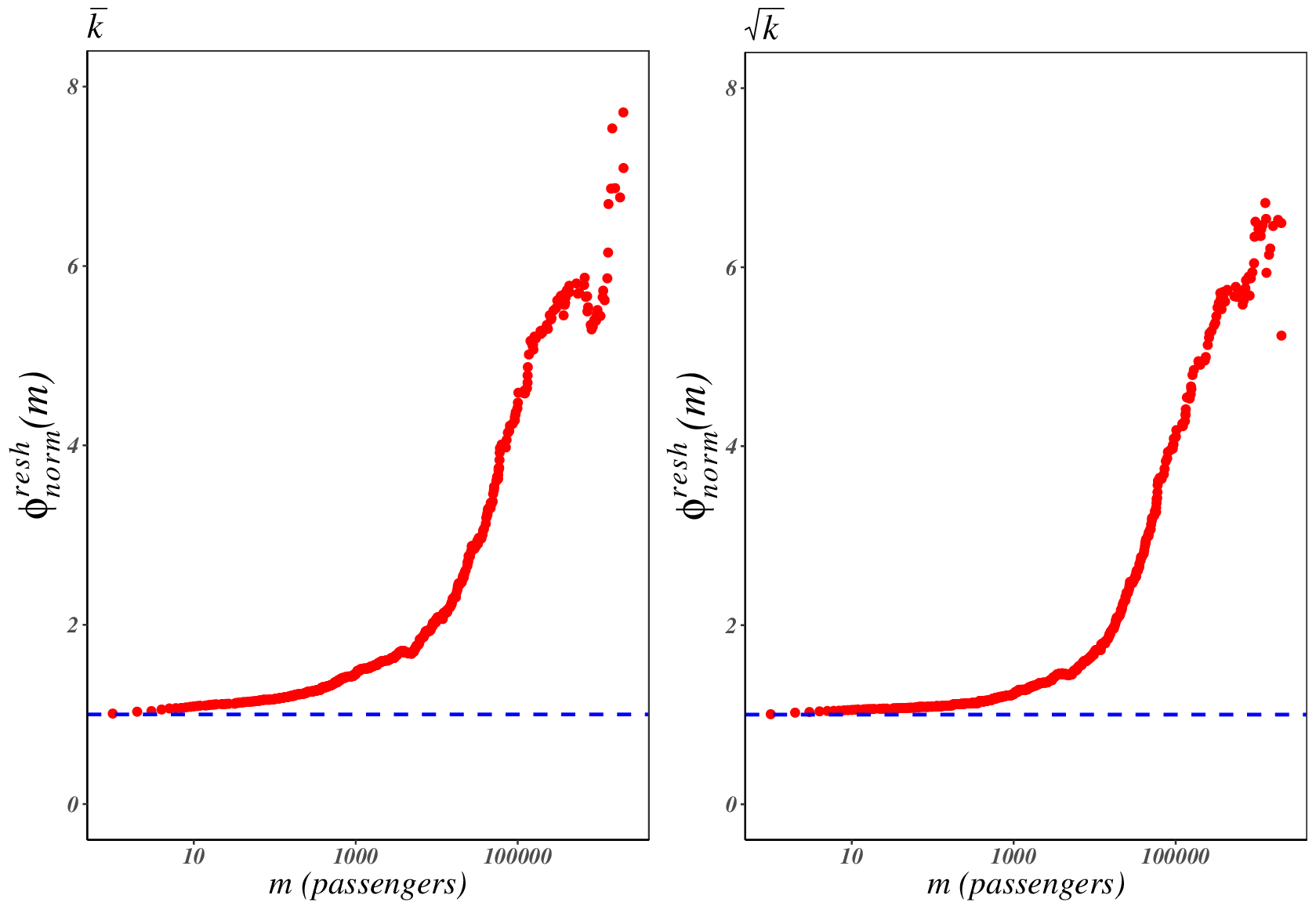}
\caption{Results for the US airports network. We compute GRCO in the case of node metadata (number of passengers leaving) by using the presented procedure of node metadata reshuffling. We use as parameters of the procedure the average degree (left) and the square root of the node degree (right).}
\label{othercase}
\end{center}
\end{figure}

It is worth noting that, asymptotically, the proposed procedure and the random shuffling should end up with somewhat equivalent distributions of the node metadata. Indeed, by iterating the procedure for a number of times which tends to infinity, we should observe reshuffled vectors displaying a degree attribute correlation which is close to that of a randomized vector, regardless of the value of $s$.
Nonetheless, since we are comparing the two cases of link rewiring and metadata reshuffling we should also consider that in the former case the number of performed rewirings is, in general, $O(M)$ where $M$ is the number of links. %Therefore, in order to be consistent in the comparison with the rewiring procedure we should perform the same amount of rewirings and reshufflings (i.e. the same amount of rewirings and iterations of the proposed procedure). 
This implies that, in practical contexts by performing $O(M)$ iterations, the proposed procedure would produce a correlation profile which differs from the random one. 
Indeed, in Figure \ref{histograms} we observe how the proposed procedure, regardless of the chosen parameter, keeps a higher correlation profile that overlaps with the random one only in its left tail.
By using this procedure we can further test the presence of rich-club ordering on different normalized ensembles, thus obtaining the results of Figure \ref{othercase}.
The obtained results still confirm the presence of rich-club ordering in the case of node metadata but they are clearly on a different scale with respect to the case of random reshuffling, as displayed in Figure \ref{airport_rich}. The results of this stress test provide further evidence for the presence of a tight core in the considered airport network.

\section{Discussion}

In this paper we discussed the generalization of the concept of rich-club ordering, considering both node structural attributes and metadata. This allowed room for the evaluation of such a phenomenon from a number of different perspectives that embed external information about nodes and that can be useful in the study of real networks. 
For instance, when studying economic networks, such as trade networks or interbanks networks, one may be interested in noticing whether the richest agents (in an economic sense) do actually form a rich-club whilst not being hubs. In other words, whether they tend to saturate their degree by connecting only to other rich-members, thus minimizing their feeder (i.e. rich-club to non rich-club) connections. The study of such feeder connections, whose endpoints are nodes outside the rich-club, i.e. nodes which can be in a certain proportion considered eligible to join the rich-club, has proved to be important in confirming the presence of rich-club ordering \cite{cinelli2018richclub} and it could provide insights for the understanding of the dynamical properties of the rich-club which are, like the growth, still largely unexplored \cite{ma2015richcores}.
Moreover, GRCO can be easily extended to the case of weighted (i.e. networks with edge metadata) and directed networks by using the right null models for these specific cases \cite{serrano2008rich,opsahl2008prominence,zlatic2009rich}.

%Moreover, the introduction of generalized rich-club ordering extends the definition of richness to external sources of data and it can be helpful in the study of many different networks. For instance, when studying economic networks, such as trade networks or interbanks networks, one may be interested in noticing whether the richest agents (in the economic notion) do actually form a rich-club also not being hubs. In other words, if they tend to saturate their degree by connecting only to other rich-members thus minimizing their feeder (i.e. rich-club to non rich-club) connections. The study of such feeder connections, whose endpoints are nodes outside the rich-club, thus  nodes which can be considered somehow eligible to join the rich-club, has proved to be important in confirming the presence of rich-club ordering \cite{cinelli2018richclub}. The study of such feeder connections also represents an interesting element to investigate in the case of different definitions of rich-club ordering and it can provide insights into the dynamical properties, like the growth, of the rich-club which are still largely unexplored \cite{ma2015richcores}.

This generalization also aims at shedding more light on the relationship that exists between topological and non-topological patterns in real networks, as well as at emphasizing the importance of node metadata. Given the current possibility to collect and store increasingly richer datasets and networks, the metadata are indeed gaining attention in Network Science and many topological phenomena such as the Friendship Paradox \cite{zuckerman2001what} (which states that your friends have, on average, more friends than you have), are now being generalized considering the presence of node characteristics \cite{eom2014generalized}. The use of such metadata has also been extended to other topological network properties, such as motifs \cite{milo2002network}, that are now enriched considering their functional aspects when examined in real networks \cite{van2012high}.

Additionally, we discussed the importance of testing rich-club ordering with the appropriate null models, which can provide us with a deeper understanding of the numerous facets of this problem. However, such an approach always implies a trade-off between what can be kept and what can be dropped regarding the network structure and its relation to the node metadata.

\section*{Acknowledgments}

The author thanks Leto Peel, Ra\`{u}l J. Mondrag{\'o}n and Antonio Iovanella for their insightful suggestions and comments.

%\bibliography{Alexandria_Cin}


\end{document}